\documentclass[twoside]{article}
\usepackage{PRIMEarxiv}

\usepackage[utf8]{inputenc} 
\usepackage[T1]{fontenc}    
\usepackage{hyperref}       
\usepackage{url}            
\usepackage{booktabs}       
\usepackage{amsfonts}       
\usepackage{nicefrac}       
\usepackage{microtype}      
\usepackage{lipsum}
\usepackage{fancyhdr}       
\usepackage{graphicx}       
\graphicspath{{media/}}     
\usepackage{setspace}

\usepackage{cite}
\usepackage{amsmath,amssymb,amsthm,mathrsfs,amsfonts,dsfont} 
\DeclareMathAlphabet{\mathpzc}{OT1}{pzc}{m}{it}
\usepackage{algorithm,algorithmic,setspace}
\usepackage{graphicx}
\usepackage{textcomp}
\usepackage{xcolor}
\usepackage{tabularx,booktabs}
\usepackage{adjustbox}
\usepackage{array,etoolbox}
\usepackage{dcolumn}
\usepackage{rotating}
\usepackage{pifont}
\usepackage{tikz}
\usepackage{comment}
\usepackage{dsfont}
\usepackage{multicol}
\usepackage{multirow}
\usepackage{mathtools}
\usepackage{stfloats}
\usepackage{lettrine}

\usepackage[nopostdot,style=super,nonumberlist, toc]{glossaries}
\usepackage{tabularx,booktabs}
\usepackage[utf8]{inputenc}
\usepackage{fourier} 
\usepackage{array}
\usepackage{makecell}

\title{QoS-aware State-Augmented Learnable Algorithm for 
Wireless Coexistence Parameter Management
\thanks{This work has been submitted to the IEEE for possible publication.  Copyright may be transferred without notice, after which this version may no longer be accessible.}
}

\author{
  Mohammad Reza Fasihi, Brian L. Mark \\
  Dept. of Electrical and Computer Engineering and Wireless Cyber Center \\
  George Mason University \\
  Fairfax, Virginia, United States\\
  \texttt{\{mfasihi4, bmark\}@gmu.edu} \\
}

\begin{document}
\maketitle

\begin{abstract}
Efficient and fair coexistence in unlicensed spectrum is essential to support heterogeneous networks such as 5G NR-U and Wi-Fi, which often contend for shared wireless resources. We introduce a general framework for
wireless Coexistence Parameter Management (CPM) based on state-augmented constrained reinforcement learning. 
We propose a novel 
algorithm, QaSAL-CPM, which incorporates state augmentation by embedding the dual variables 
in the constrained optimization formulation
directly into the agent's observation space. This method enables the agent to respond to constraint violations in real time while continuing to optimize a primary performance objective. 
Through extensive simulations of 5G NR-U and Wi-Fi coexistence scenarios, we show that QaSAL-CPM achieves reliable QoS compliance and improved policy robustness across various transmitter densities compared to previous approaches. The proposed
framework offers a scalable and adaptive solution for real-time coexistence optimization in next-generation wireless networks.
\end{abstract}

\keywords{
5G NR-U \and Wi-Fi \and coexistence \and unlicensed spectrum \and 
quality of service \and medium access delay \and fairness}

\section{Introduction}

\lettrine[findent=2pt]{\textbf{T}}{he rapid growth in wireless connectivity} demands driven by applications ranging from mobile broadband to the Internet of Things (IoT) has led to increased reliance on unlicensed spectrum. These frequency bands are shared by multiple Radio Access Technologies (RATs), such as 5G New Radio in Unlicensed Spectrum (NR-U) and Wi-Fi, creating complex interference dynamics that, if left unmanaged, can significantly degrade network performance. The key challenge lies in ensuring high performance while maintaining fairness among the coexisting technologies sharing the same spectrum~\cite{Sathya:2020, Saha:2021, Hirzallah:2021}.

The coexistence of 5G NR-U and Wi-Fi introduces unique difficulties due to their distinct medium access protocols. While NR-U relies on Listen-Before-Talk (LBT), Wi-Fi uses Carrier Sense Multiple Access with Collision Avoidance (CSMA/CA). These different
protocols should coordinate effectively to minimize inter-network collisions and achieve fair spectrum utilization. The situation is further complicated by divergent Quality of Service (QoS) requirements and traffic behaviors. For instance, 5G NR-U often supports latency-sensitive, high-priority services that may conflict with Wi-Fi’s best-effort traffic model. Moreover, the dynamic nature of unlicensed spectrum usage—due to fluctuating traffic loads, mobility, and environmental changes—demands adaptive and robust coexistence strategies~\cite{Muhammad:2020}. Without such strategies, the performance of both technologies can deteriorate due to increased contention, latency, and unfair access. Managing contention and enhancing performance under these conditions can be formulated as a QoS-aware network utility maximization problem, in which the tuning of Medium Access Control (MAC) parameters (e.g., contention window sizes), which we refer to as \textit{Coexistence Parameter Management} (CPM), is key.

Multi-objective reinforcement learning (RL) has been applied to this problem by defining rewards as weighted combinations of performance metrics such as delay and fairness~\cite{Mannor:2004, Moffaert:2013}. However, this approach requires manually selecting and tuning weights, which introduces substantial computational overhead and does not guarantee strict adherence to QoS requirements. In contrast, constrained RL explicitly incorporates QoS constraints into the learning process~\cite{Bhatnagar:2012}. Constrained RL formulates the problem using \textit{Lagrangian} relaxation, where a single objective function is optimized while constraints are enforced through dynamic dual variables. This method automatically adjusts 
the Lagrange multipliers, simplifying the design and enabling better constraint satisfaction over time. Still, dual-based methods can struggle in highly dynamic environments due to difficulties in tuning penalty terms and ensuring stable convergence~\cite{Fullana:2021,NaderiAlizadeh:2022}.

In this paper, we propose a general framework for CPM based on a \textit{state-augmented} RL formulation, which was developed
recently in~\cite{Fullana:2024,Uslu:2025} to overcome the drawbacks of the dual-based methods. In state-augmented RL, the dual variables are embedded
directly into the agent’s observation space, enabling real-time responsiveness to constraint violations and reducing 
reliance on hand-tuned penalties. Building upon our earlier works~\cite{Fasihi:2024, Fasihi:2025}, which introduced traffic-aware RL and the application of state-augmented RL to the wireless coexistence problem, we present a new algorithm, QaSAL-CPM, which integrates state augmentation with a Double Deep Q-Network (DDQN) and experience replay. QaSAL-CPM achieves real-time QoS guarantees for high-priority traffic while maintaining fairness across all traffic classes in a wireless coexistence scenario.
We implemented a Python-based simulation environment that models the MAC-layer behavior of 5G NR-U and Wi-Fi under saturated traffic conditions to compare the performance of QaSAL-CPM to that of multi-objective RL and
constrained RL based on conventional primal-dual methods.  The proposed QaSAL-CPM provides a framework for
significantly improved QoS compliance and policy robustness across diverse wireless coexistence
scenarios.

The rest of the paper is organized as follows: Section~\ref{sec:RelatedWorks} reviews related work on general coexistence strategies, the application of RL to NR-U/Wi-Fi coexistence, and constrained RL techniques. Section~\ref{sec:SystemModel} presents the system model and formally defines the CPM problem. In Section~\ref{sec:MORL-CPM}, we develop a 
multi-objective RL approach for CPM, which extends our prior work in~\cite{Fasihi:2024}.
Section~\ref{sec:CRL-CPM} introduces a constrained RL formulation, discusses a primal-dual method for solving it, and presents the state-augmented RL framework. This motivates the design of our proposed algorithm, QaSAL-CPM, which is detailed in Section~\ref{sec:QaSAL-CPM} along with its deep~RL implementation. Section~\ref{sec:QaSAL-CPM-Coexistence} applies the CPM framework to the coexistence of 5G NR-U and Wi-Fi in unlicensed spectrum, discusses the performance metrics of interest,
the simulation setup, and the numerical results. The paper is concluded in Section~\ref{sec:Conclusion}.
\section{Related Works}
\label{sec:RelatedWorks}


\subsection{General Coexistence Studies}

Coexistence of 5G NR-U/LAA and WiFi has been studied extensively in the recent literature. For example, \cite{Sathya:2020} and \cite{Saha:2021} provide comprehensive overviews of
the state-of-the-art research on coexistence of different technologies in sub-7 GHz unlicensed spectrum. A detailed overview of the new 5G NR-U specifications and its coexistence with Wi-Fi in unlicensed bands, focusing on key challenges in PHY, MAC, and higher-layer operations is presented in~\cite{Hirzallah:2021}. A comprehensive survey on the opportunities and challenges associated with unlicensed operations in the 6~GHz band, focusing on Wi-Fi and 5G NR-U coexistence, can be found in~\cite{Naik:2020}.
In~\cite{Farahiyah:2024}, the coexistence of licensed fixed service systems with unlicensed Radio Local Area Networks  and 5G NR-U in the 6~GHz band, is investigated, focusing on potential 5G NR-U and Wi-Fi interference risks.



\subsection{Bandit-Based Reinforcement Learning}

The \textit{Multi-Armed Bandit} (MAB) problem is a fundamental reinforcement learning framework
that models decision-making under uncertainty, where an agent must choose among multiple \textit{arms} to maximize cumulative rewards over time. In each round, the agent selects an action, observes a reward, and updates its strategy to balance exploration (gathering new information) and exploitation (leveraging past knowledge). In the context of coexistence management for unlicensed spectrum, MAB algorithms provide an adaptive mechanism for optimizing channel access, power control, and contention window adjustments in dynamic environments where 5G NR-U and Wi-Fi compete for shared resources.  

In~\cite{Shi:2020}, an MAB-based online learning distributed channel selection algorithm is
proposed to enable NR-U users to make optimal channel selections without requiring complete environmental knowledge. An AI-driven framework is introduced in~\cite{Hirzallah:2021_2} for adaptive sensing threshold selection in shared spectrum environments by leveraging a clustering-based multi-armed bandit algorithm, ensuring efficient coexistence among Wi-Fi, LTE-LAA, and 5G NR-U systems. A novel framework for fair and efficient NR-U coexistence in shared spectrum environments is presented in~\cite{Bajracharya:2023}, which formulates the channel access problem as a multi-objective MAB that 
jointly optimizes spectrum efficiency and fairness.

\subsection{State-Driven Reinforcement Learning}

MAB algorithms operate under a context-free learning paradigm, where decisions are made without considering past states or future consequences. In real-world coexistence scenarios, however, the wireless environment is highly dynamic—interference patterns, traffic loads, and network conditions fluctuate over time, requiring an approach that can learn and adapt based on these changes. State-driven RL methods address this limitation of MAB-based approaches by incorporating \textit{state transitions} and \textit{delayed rewards}, which facilitates long-term optimization of coexistence strategies. 

\begin{figure*}
\centering
	\includegraphics[scale=0.25]{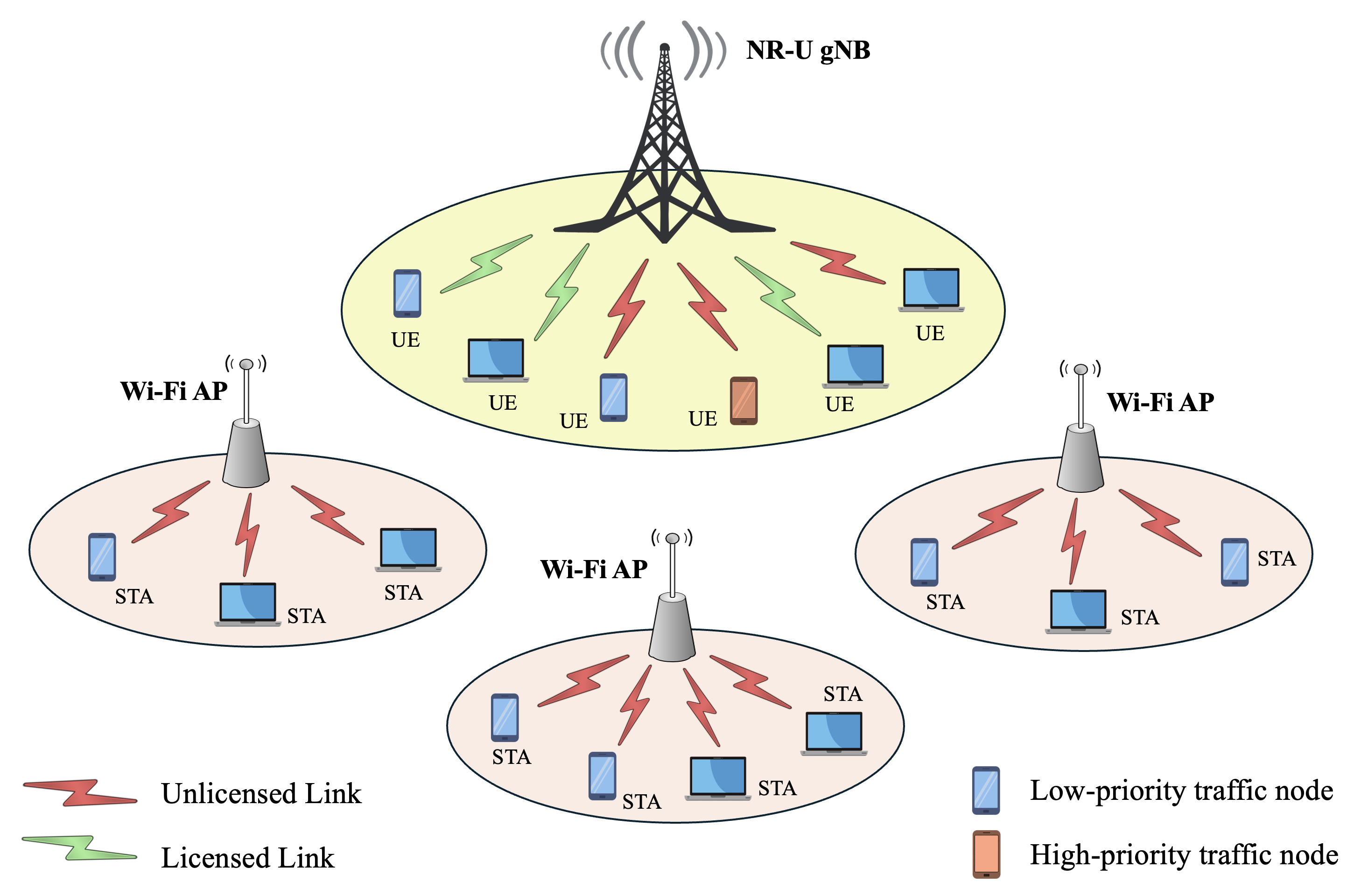}
    \caption{A 5G NR-U/Wi-Fi coexistence in an unlicensed spectrum with multi-priority traffic transmitter.}
	\label{fig:CoexistenceEnvironment}
\end{figure*}

In~\cite{Wang:2021}, an RL-based approach is proposed for joint allocation of transmission opportunities and spectral resources in 5G NR-U and Wi-Fi coexistence while ensuring fair spectrum sharing.  In~\cite{Zeinali:2023}, a joint base station and resource allocation framework based on deep~RL is presented for a heterogeneous network with vehicle-to-everything users. 
Deep RL-based approaches for energy detection threshold optimization in NR-U and Wi-Fi coexistence systems, focusing on downlink Ultra-Reliable Low-Latency Communications (URLLC),
are proposed in~\cite{Liu:2023}. 
In~\cite{Zou:2023}, the challenge of fair coexistence between 5G NR and Wi-Fi while accommodating URLLC traffic is addressed through
a mixed priority scheduling mechanism that aims to balance latency constraints and data size. 
The previous RL-based approaches do not explicitly address QoS constraints within the wireless coexistence context, which is
the main contribution of our work.

\section{System Model and Problem Formulation}
\label{sec:SystemModel}

\subsection{System Model}

We consider a 5G NR-U network coexisting with Wi-Fi networks on an unlicensed frequency band as illustrated in Fig.~\ref{fig:CoexistenceEnvironment}. Transmitters of both networks contend with each other to access the channel for downlink transmissions. We focus on a worst-case saturated scenario, where transmitters in both networks always have packets to transmit. To access the channel, the Wi-Fi network adopts the IEEE 802.11 enhanced distributed channel access (EDCA) protocol, which uses binary exponential backoff with a random initial backoff value. EDCA defines four access classes, each of which has a different initial backoff window size, maximum backoff stages, and maximum transmission opportunity (TXOP) duration. 

NR-U gNB transmitters, on the other hand, employ the LBT mechanism, which is consistent with Wi-Fi's EDCA,
and supports four access priority classes that correspond to the four access classes in EDCA~\cite{ETSI_TS_137213}. For the PHY layer, gNBs have flexible numerologies and time slot scheduling in which each NR-U transmission can only begin at the next slot boundary. If the channel is found idle by a gNB, it transmits a reservation signal (RS) until the next slot boundary to prevent other transmitters from occupying the channel. For simplicity, we assume that each NR-U transmitter has the same channel occupation times for successful transmissions and collisions as Wi-Fi.  

Without loss of generality, we assume two types of transmitters for each network: 1) high-priority traffic transmitters with strict delay requirements and 2) low-priority traffic transmitters with less stringent delay constraints. Throughout this work, we denote the transmitters with high-priority and low-priority traffic by \textit{PC1} (priority class 1) and \textit{PC3} (priority class 3), respectively. The LBT protocol can harm the latency performance of PC1 traffic such as URLLC packets, especially when contending for channel access with PC3 traffic~\cite{Le:2021}. Additionally, the LBT protocol in NR-U may cause additional transmission delay compared to 5G~NR in licensed spectrum due to the unpredictability of transmission opportunities. Therefore, the delay performance of PC1 traffic may 
not always meet the requirements. 



\begin{figure*}
\centering
	\includegraphics[scale=0.30]{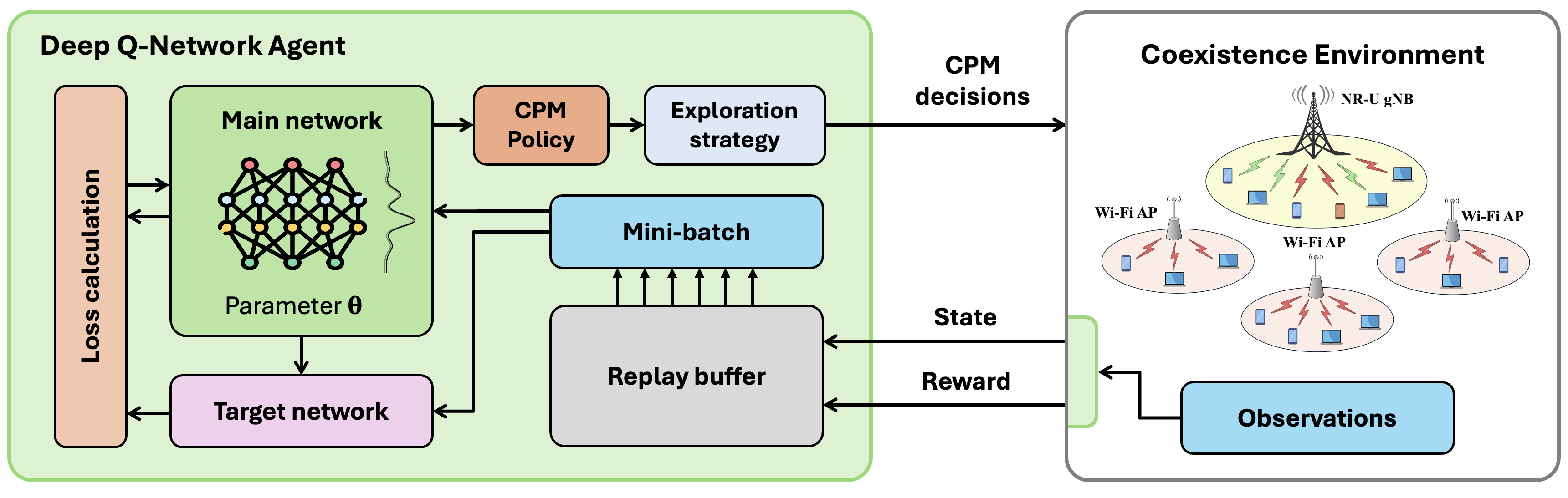}
    \caption{Illustration of parameterized optimization of the CPM policy through Deep Q-Network (DQN).}
	\label{fig:CPM_Problem}
\end{figure*}

Effective coexistence management between 5G NR-U and Wi-Fi in unlicensed spectrum requires adaptive parameter tuning to balance fair spectrum access and QoS performance (e.g., latency, efficiency, and throughput) for PC1 traffic transmitters. Examples of key coexistence parameters include the Contention Window (CW) size and the Arbitration Inter-Frame Space (AIFS) duration, which directly impact how transmitters of both networks compete for channel access.
CPM refers to the dynamic adjustment of these parameters based on real-time network conditions. The objective of CPM is to ensure the QoS requirements for high-priority traffic while maintaining fair and efficient coexistence. Fig.~\ref{fig:CPM_Problem} illustrates the parameterized optimization of the CPM policy through a Double Deep Q-Network (DDQN), 
which will be discussed further in Section~\ref{sec:QaSAL-CPM}.

\subsection{Problem Formulation}

Let $\mathpzc{S}\subset\mathds{R}^n$ represent a compact set of \textit{coexistence environment states} capturing the status of both the 5G NR-U and Wi-Fi networks. Given the state $\mathbf{S}_t\in\mathpzc{S}$, let $\mathbf{a}(\mathbf{S}_t)$ denote the vector of CPM decisions across both networks, where $\mathbf{a}:\mathpzc{S}\rightarrow\mathds{R}^a$ is the CPM function. The possible states and CPM decisions are described by a Markov Decision Process (MDP) with transition probability $p\left(\mathbf{S}_{t+1}|\mathbf{S}_t, \mathbf{a}(\mathbf{S}_t)\right)$. The agent sequentially makes CPM decisions over discrete time steps $t\in\mathds{N}\cup\{0\}$ based on a policy $\pi$, and observes the performance vector $\mathbf{f}(\mathbf{S}_t, \mathbf{a}(\mathbf{S}_t))\in\mathds{R}^m$ which captures various performance metrics of both networks, such as delay, collision rate, airtime efficiency, fairness, number of successful transmissions, channel utilization ratio, etc. 

As with any MDP, the primary concern is the network performance the agent accumulates over
an infinite horizon, which is represented by the value function 
\begin{equation}
    V_i(\pi)\triangleq\lim_{T\rightarrow\infty}\frac{1}{T}\mathds{E}_{\left(\mathbf{S},\mathbf{a}(\mathbf{S})\right)\sim\pi}\left[\sum_{t=0}^{T} f_i(\mathbf{S}_t, \mathbf{a}(\mathbf{S}_t))\right],
    \label{eq:ValueFunction}
\end{equation}
where $i=0,\ldots,m-1$. In practice, because the probability distribution of the environment is unknown, the objective is derived based on the \textit{ergodic average} network performance
\begin{equation}
    \Tilde{V}_i(\pi)=\frac{1}{T}\sum_{t=0}^{T-1} f_i(\mathbf{S}_t, \mathbf{a}(\mathbf{S}_t)).
    \label{eq:ErgodicValueFunction}
\end{equation}


In the coexistence of 5G NR-U and Wi-Fi, multiple conflicting objectives naturally arise due to the need to balance different network performance requirements. Each network operates under distinct access mechanisms and traffic demands, leading to trade-offs between key performance metrics. Some transmitters may require strict QoS guarantees, while others prioritize overall network efficiency or fairness in resource allocation. Optimizing one aspect of the system often comes at the expense of another, making it challenging to achieve an optimal balance through single-objective optimization. This makes CPM a \emph{multi-objective} optimization problem:
\begin{equation}
    \max{\Tilde{\mathbf{V}}(\pi)}=\max{\left\{\Tilde{V}_0(\pi), \Tilde{V}_1(\pi), \dots, \Tilde{V}_{m-1}(\pi)\right\}} ,
    \label{eq:MO-CPM}
\end{equation}
in which different performance criteria must be optimized together rather than in isolation.

\section{Multi-Objective Optimization for CPM}
\label{sec:MORL-CPM}

To solve~\eqref{eq:MO-CPM}, a structured approach is needed to systematically weigh the different objectives and guide the learning process toward solutions that achieve the best possible trade-offs. A common approach to addressing this challenge is \textit{scalarization}, which applies a function that combines all objectives into a single objective function. This reformulation allows the problem to be treated as a standard single-objective MDP, enabling the agent to learn a policy that maximizes the scalarized objective. 
A common scalarization approach is \textit{linear scalarization}, which involves a weight vector $\boldsymbol{w}=\{w_0, w_1, \dots, w_{m-1}\}$, $w_i\in[0,1]$, $\sum_{i=0}^{m-1}{w_i}=1$. Weight~$w_i$ indicates the
relative importance of objective $f_i(\mathbf{S}_t, \mathbf{a}(\mathbf{S}_t))$ in a weighted sum of the objectives:
\begin{equation}
    f_{\text{scalar}}(\mathbf{S}_t, \mathbf{a}(\mathbf{S}_t))=\sum_{i=0}^{m-1}{w_i f_{i}(\mathbf{S}_t, \mathbf{a}(\mathbf{S}_t))}.   
    \label{eq:Linear-Scalarized-Objective}
\end{equation}

Optimization of $f_{\text{scalar}}(\mathbf{S}_t, \mathbf{a}(\mathbf{S}_t))$ in (\ref{eq:Linear-Scalarized-Objective}) involves an infinite-dimensional search over the space of CPM decisions $\mathbf{a}(\mathbf{S})$ for any given network state $\mathbf{S}$, making direct optimization impractical. To address this, a parameterized approach for the CPM policy can be adopted by replacing $\mathbf{a}(\mathbf{S})$ with $\mathbf{a}(\mathbf{S};\boldsymbol{\theta})$, where $\boldsymbol{\theta}\in\boldsymbol{\Theta}$ and $\boldsymbol{\Theta}$ denotes a finite-dimensional set of parameters, and maximization is performed iteratively over the set $\boldsymbol{\theta}$. This leads to the \textit{parameterized multi-objective CPM} problem 
\begin{equation}
    \max_{\boldsymbol{\theta}\in\boldsymbol{\Theta}}{\frac{1}{T}\sum_{t=0}^{T-1}{f_{\text{scalar}}(\mathbf{S}_t, \mathbf{a}(\mathbf{S}_t;\boldsymbol{\theta}))}},
    \label{eq:Parameterized-Linear-Scalarized-CPM}
\end{equation}
which can be solved through multi-objective RL algorithms. Although linear scalarization is computationally efficient and easy to implement, it assumes that the trade-offs between objectives can be accurately captured by the predefined weights in (\ref{eq:Linear-Scalarized-Objective}). In practice, selecting appropriate weight values can be challenging, especially when objectives exhibit nonlinear dependencies. More advanced multi-objective RL approaches attempt to overcome this limitation by dynamically adjusting the weighting scheme or learning a set of Pareto-optimal policies.


\section{Constrained Reinforcement Learning for CPM}
\label{sec:CRL-CPM}

As an alternative to multi-objective RL, the QoS requirements can be explicitly incorporated into a constrained RL problem. To formulate the CPM problem within constrained RL, we designate the first objective function in (\ref{eq:MO-CPM}), $f_0(\mathbf{S}_t, \mathbf{a}(\mathbf{S}_t))$, as the \textit{primary} objective to maximize, while treating the remaining objectives as \textit{QoS constraints} that must be satisfied. The goal of the CPM problem is to determine the optimal CPM decision $\mathbf{a}(\mathbf{S}_t)$ vector for any given network state $\mathbf{S}_t\in\mathpzc{S}$, such that the primary objective is optimized while ensuring compliance with QoS constraints. Accordingly, the generic \textit{parameterized constrained CPM} problem is defined as
\begin{subequations}
    \begin{align}
        \max_{\boldsymbol{\theta}\in\boldsymbol{\Theta}} \quad &
            \frac{1}{T}\sum_{t=0}^{T-1} f_0(\mathbf{S}_t, \mathbf{a}(\mathbf{S}_t;\boldsymbol{\theta})),
            \label{eq:ParameterizedCPM:Objective} \\
        \text{s.t.} \quad &
            \frac{1}{T}\sum_{t=0}^{T-1} f_i(\mathbf{S}_t, \mathbf{a}(\mathbf{S}_t;\boldsymbol{\theta}))
            \geq \text{c}_i, \quad i=1,\dots,m-1 ,
            \label{eq:ParameterizedCPM:Constraints}
    \end{align}
    \label{eq:Parameterized-CRL-CPM}
\end{subequations} 
where the constant $\text{c}_i\in\mathds{R}$ represents the threshold value for the $i$-th objective function. In this paper,
we develop a learning algorithm to solve~(\ref{eq:Parameterized-CRL-CPM}) for any given coexistence environment state $\mathbf{S}_t\in\mathpzc{S}$.

\subsection{Primal-Dual Approach for Constrained CPM}

A customary approach to solve~(\ref{eq:Parameterized-CRL-CPM}) is to consider a penalized version in the \textit{Lagrangian dual} domain. Formally, we introduce dual variables $\boldsymbol{\lambda}\in\mathds{R}_+^c$ associated with the constraints in (\ref{eq:ParameterizedCPM:Constraints}) and define the Lagrangian
\begin{equation}
        \mathpzc{L}(\boldsymbol{\lambda};\boldsymbol{\theta}) = \frac{1}{T}\sum_{t=0}^{T-1} f_0(\mathbf{S}_t, \mathbf{a}(\mathbf{S}_t;\boldsymbol{\theta})) + \sum_{i=1}^{m-1}{\lambda_i\left(\left(\frac{1}{T}\sum_{t=0}^{T-1} f_i(\mathbf{S}_t, \mathbf{a}(\mathbf{S}_t;\boldsymbol{\theta}))\right) - \text{c}_i\right)}.
    \label{eq:ParameterizedLagrangian}
\end{equation}

The Lagrangian in (\ref{eq:ParameterizedLagrangian}) should be maximized over $\boldsymbol{\theta}$, while subsequently minimizing over the dual variables $\boldsymbol{\lambda}$, i.e.,
\begin{equation}
    \min_{\boldsymbol{\lambda}\in\mathds{R}_+^c} \quad \max_{\boldsymbol{\theta\in\Theta}} \quad \mathpzc{L}(\boldsymbol{\lambda};\boldsymbol{\theta}).
    \label{eq:MinMaxProblem}
\end{equation}

The advantage of replacing the objective in (\ref{eq:Parameterized-CRL-CPM}) with
the Lagrangian in~(\ref{eq:ParameterizedLagrangian}) is that the latter can be optimized using any parameterized learning framework, such as standard RL algorithms. One limitation of (\ref{eq:MinMaxProblem}) is the ambiguity in determining suitable values for the dual variables. The optimal choice for $\boldsymbol{\lambda}$ depends on the transition probability $p(\mathbf{S}_{t+1}|\mathbf{S}_t,\mathbf{a}(\mathbf{S}_t;\boldsymbol{\theta}))$, which is typically unknown. This challenge can be circumvented by dynamically adjusting the dual variables $\boldsymbol{\lambda}$. To accomplish this, we define an iteration index $k\in \left\{0,1,\dots,\left\lfloor T/T_0 \right\rfloor-1\right\}$, where $T_0$ is the duration of one epoch, which is the number of time steps between successive updates of the model parameters. Also, we introduce the learning rates $\eta_{\lambda_i}\in\mathds{R}_+$ corresponding to each $\lambda_i$. The model parameters $\boldsymbol{\theta}_k$ and dual variables $\lambda_{i,k}$ are updated iteratively according to equations (\ref{eq:LambdaUpdateIterationPrimalDual}) and (\ref{eq:LambdaUpdateIteration}). 

During the $k$-th epoch, the agent collects samples over a window of length $T_0$ and uses them to compute the constraint violation. The primal update in (\ref{eq:LambdaUpdateIterationPrimalDual}) seeks to maximize the Lagrangian, which includes the main objective $f_0$ and a penalty term weighted by the current dual variable $\lambda_{i,k}$ and the degree of constraint violations. The dual update in (\ref{eq:LambdaUpdateIteration}) adjusts each $\lambda_{i,k}$ based on the extent to which its corresponding constraint $\text{c}_i$ is violated. If the objective $f_i$ exceeds $\text{c}_i$, the dual variable increases proportionally to the violation, encouraging the agent to reduce the violation in future updates. The $[\cdot]^+$ operator ensures that the dual variables remain non-negative. The update rate $\eta_{\lambda_i}$ can be tuned individually for each constraint to control how aggressively violations are penalized.



It has been shown that the resulting algorithm is both feasible and near-optimal when run for a sufficiently large number of time steps. This is particularly noteworthy because, in standard primal-dual methods, the learned CPM policies are not guaranteed to yield a feasible set of decisions. In contrast, the CPM algorithm defined by equations (\ref{eq:LambdaUpdateIterationPrimalDual}) and (\ref{eq:LambdaUpdateIteration}) ensures feasibility and 
near-optimality. Moreover, this guarantee holds even when the underlying optimization problem is non-convex~\cite{Fullana:2021}. 

\begin{equation}
    \boldsymbol{\theta}_{k} = \arg\max_{\boldsymbol{\theta}\in\boldsymbol{\Theta}}\left[
    \frac{1}{T_0}\sum_{t=kT_0}^{(k+1)T_0-1} f_0(\mathbf{S}_t, \mathbf{a}(\mathbf{S}_t;\boldsymbol{\theta})) 
    + \sum_{i=1}^{m-1}\lambda_{i,k} \left(\frac{1}{T_0}\sum_{t=kT_0}^{(k+1)T_0-1}\left( f_i(\mathbf{S}_t, \mathbf{a}(\mathbf{S}_t;\boldsymbol{\theta}))-\text{c}_i\right)\right)\right],
\label{eq:LambdaUpdateIterationPrimalDual}
\end{equation}
\begin{equation}
    \lambda_{i,k+1} = \left[\lambda_{i,k}-\frac{\eta_{\lambda_i}}{T_0}\sum_{t=kT_0}^{(k+1)T_0-1} \left(f_i(\mathbf{S}_t, \mathbf{a}(\mathbf{S}_t;\boldsymbol{\theta}_k))-\text{c}_i\right)\right]^+, \quad i=1,\dots,m-1.
\label{eq:LambdaUpdateIteration}
\end{equation}

One of the important limitations of the primal-dual approach is that it only guarantees convergence to a feasible and near-optimal solution in the long run, i.e., as the total operation time $T$ becomes very large, theoretically approaching infinity. In practice, this means we cannot simply stop the algorithm after a finite number of iterations and claim that the solution it has reached is close to optimal. The performance and feasibility guarantees only hold in the limit, which may not be practical for systems that need to make decisions in real-time or within limited time frames.

Another challenge with this method is that optimizing the Lagrangian function at each iteration requires information about future network states. For example, when the algorithm reaches the beginning of the $k$-th epoch (at time step $t=kT_0$), the optimization process needs access to all network states from $t=kT_0$ to $t=(k+1)T_0-1$. However, in an online or real-time setting, this future information is not available, making it impossible to perform the optimization exactly as required. While this may be manageable in offline training where future states can be simulated or assumed, it poses a serious limitation during actual deployment. 

\subsection{State-Augmented Constrained Reinforcement Learning}
\label{sec:StateAugmentedCRL-CPM}

To address the aforementioned challenges, the \textit{state-augmentation} approach provides a more practical and scalable alternative~\cite{Fullana:2021, Fullana:2024, NaderiAlizadeh:2022}. The key idea of state-augmentation is to treat constraint satisfaction as a dynamic component of the agent's environment, which evolves in response to constraint violations through dual variable dynamics. By augmenting the state with dual variables, the agent learns policies that are constraint-aware and adaptable, leading to feasible, near-optimal solutions that traditional methods cannot guarantee. 

State-augmentation approach reformulates the constrained~RL problem by incorporating the dual variables directly into the agent's state space. This simple yet powerful idea transforms the constrained optimization problem into a \textit{state augmented} MDP, where the augmented state includes both the coexistence network's state and the current values of the dual variables. By doing so, the agent is able to learn a single, adaptive policy that responds dynamically to changes in the dual variables, eliminating the need to optimize over multiple parameter sets or forecast future network states. Moreover, as discussed earlier, when dual variables evolve according to gradient-based updates, the learned policy remains feasible and near-optimal under general non-convex settings.

The state-augmentation approach offers several advantages over the classical Lagrangian-based method, where constraints are typically integrated into the reward function using fixed or static Lagrange multipliers. By embedding the dual variables directly into the agent’s state space and updating them dynamically, the agent is able to adaptively learn how to satisfy constraints over time. This dynamic adaptation is particularly beneficial in non-stationary or evolving environments, where fixed multipliers may fail to capture changing system dynamics.
The advantage of state-augmentation approach becomes particularly significant in the context of CPM, where the environment is inherently dynamic and characterized by complex interactions between multiple heterogeneous transmitters where QoS requirements can vary rapidly.
By continuously aligning coexistence parameter adjustments with the real-time status of QoS constraints, state augmentation helps ensure that critical requirements—such as bounded medium access delay for high-priority traffic—are upheld even in highly variable environments. This makes it particularly well-suited for managing the multi-objective and time-sensitive nature of CPM.

\section{Proposed Algorithm: QaSAL-CPM}
\label{sec:QaSAL-CPM}

In light of the state-augmentation approach, we propose the \textit{QoS-aware State-Augmented Learnable algorithm} for the CPM problem (QaSAL-CPM). Let us consider state $\mathbf{S}_t$ at time step $t$ of the $k$-th epoch. Augmentation of the dual variables $\boldsymbol{\lambda}_k$ into the state space results in a augmented state $\Tilde{\mathbf{S}}_t=(\mathbf{S}_t, \boldsymbol{\lambda}_k$). To be distinguished from previously introduced parameterization in Section~\ref{sec:CRL-CPM}, we introduce an alternative parameterization for the CPM policy, in which the CPM decisions are represented via the parameterization $\mathbf{a}(\mathbf{\Tilde{S}};  \boldsymbol{\Tilde{\theta}})$, where $\boldsymbol{\Tilde{\theta}}\in\boldsymbol{\Tilde{\Theta}}$ denotes the set of parameters of the state-augmented CPM policy. Then, we define the augmented version of the Lagrangian in (\ref{eq:ParameterizedLagrangian}) as
\begin{equation}
    \mathpzc{L}(\boldsymbol{\lambda};\boldsymbol{\Tilde{\theta}}) = \frac{1}{T}\sum_{t=0}^{T-1} f_0(\mathbf{\Tilde{S}}_t, \mathbf{a}(\mathbf{\Tilde{S}}_t;\boldsymbol{\Tilde{\theta}})) + \sum_{i=1}^{m-1}{\lambda_i\left(\left(\frac{1}{T}\sum_{t=0}^{T-1} f_i(\mathbf{\Tilde{S}}_t, \mathbf{a}(\mathbf{\Tilde{S}}_t;\boldsymbol{\Tilde{\theta}}))\right) - \text{c}_i\right)}.
    \label{eq:ParameterizedAugmentedLagrangian}
\end{equation}

\begin{figure*}
\centering
\includegraphics[scale=0.32]{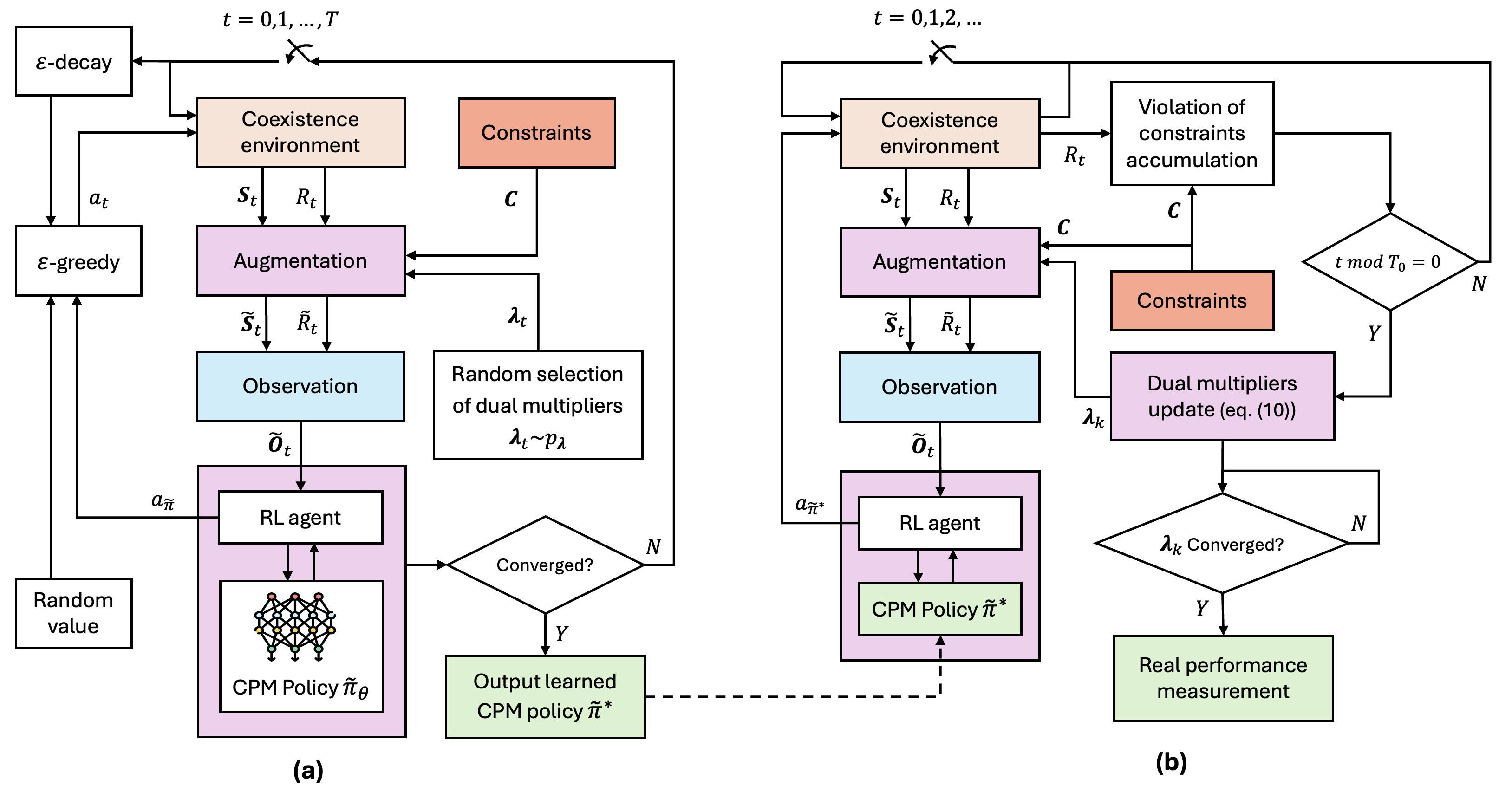}
    \caption{Flowchart of QaSAL-CPM: a) Training phase b) Execution phase.}
    \label{fig:QaSAL_Algorithm}
\end{figure*}

Considering a probability distribution $p_{\boldsymbol{\lambda}}$ for the dual variables, we define the optimal \textit{state-augmented CPM} policy which maximizes the expected augmented Lagrangian over this distribution as
\begin{equation}
    \boldsymbol{\Tilde{\theta}}^* = \arg\max_{\boldsymbol{\Tilde{\theta}}\in\boldsymbol{\Tilde{\Theta}}} ~\mathds{E}_{\boldsymbol{\lambda}\sim p_{\boldsymbol{\lambda}}}{\left\{\mathpzc{L}(\boldsymbol{\lambda};\boldsymbol{\Tilde{\theta}})\right\}}.
    \label{eq:MinMaxAugmentedProblem}
\end{equation}
Using the augmented policy parameterized by (\ref{eq:MinMaxAugmentedProblem}), we can obtain the CPM decisions $\mathbf{a}(\mathbf{\Tilde{S}}; \boldsymbol{\Tilde{\theta}})$ which maximize the Lagrangian corresponding to the dual variables $\boldsymbol{\lambda}$ at each iteration $k$. The dual variable update equation in (\ref{eq:LambdaUpdateIteration}) should also be replaced with its augmented version:
\begin{equation}
    \lambda_{i,k+1} = \left[\lambda_{i,k}-\frac{\eta_{\lambda_i}}{T_0}\sum_{t=kT_0}^{(k+1)T_0-1} \left(f_i(\mathbf{\Tilde{S}}_t, \mathbf{a}(\mathbf{\Tilde{S}}_t;\boldsymbol{\Tilde{\theta}}_k^*))-\text{c}_i\right)\right]^+,
    \label{eq:AugmentedLambdaUpdateIteration}
\end{equation}
where $i=1,\dots,m-1$. 

To ensure the effectiveness of this approach, the chosen parameterization must possess sufficient expressive power. Specifically, the CPM decisions generated by (\ref{eq:MinMaxAugmentedProblem}) should be able to closely approximate those that would be obtained by solving the original constrained optimization problem iteratively for each $\lambda$, where the dependence of $\boldsymbol{\theta}_k$ and $\boldsymbol{\lambda}_k$ is made explicit, i.e., (\ref{eq:LambdaUpdateIterationPrimalDual}) and (\ref{eq:LambdaUpdateIteration}). This requirement places the chosen parameterization within a class known as \textit{near-universal parameterizations}, which are capable of approximating the optimal mapping from dual variables to policy decisions with high fidelity. Examples of such expressive function approximators include neural networks and radial basis functions. These parameterizations are particularly effective in the constrained~RL setting where the policy must adapt dynamically to changes in dual variables. To optimize such policies, standard RL algorithms, such as value-based methods or policy-gradient-based approaches, can be employed. 

While these methods do not guarantee convergence to a globally optimal solution, they have demonstrated strong empirical performance and often converge to policies with minimal sub-optimality, especially when combined with expressive function approximators~\cite{Fullana:2024}. 
An important point is that the optimal model parameters $\boldsymbol{\Tilde{\theta}}^*$ are directly utilized in (\ref{eq:AugmentedLambdaUpdateIteration}), which mitigates the challenge posed in~(\ref{eq:LambdaUpdateIterationPrimalDual}), where computing and storing the optimal set of parameters for every possible configuration of dual variables would be required. This was, in fact, the primary motivation for adopting the state-augmented parameterization approach. 


\subsection{Deep Reinforcement Learning for QaSAL-CPM}
\label{sec:DRL-QASAL-CPM}

A flowchart depicting the QaSAL-CPM framework is given in Fig.~\ref{fig:QaSAL_Algorithm}.
We apply a Double Deep Q-Network (DDQN), which
decouples action selection and action evaluation during target computation by using the main Q-network to select the action and the target network to evaluate its value. This modification results in more accurate value estimates and improved learning stability, particularly in complex, dynamic environments. The training phase begins by initializing the model and target network parameters, as well as an empty experience replay memory. For each training episode, the environment is initialized, and a set of dual variables representing QoS constraints is sampled. These dual variables are then concatenated with the environment state to form an augmented state. The agent uses this augmented state to select a coexistence action that balances primary performance objectives with constraint satisfaction. After executing the action, the environment provides a reward that incorporates both the primary objective and penalties for constraint violations, and the agent transitions to a new augmented state. Each transition is stored in the replay memory for later training.

During training, the agent samples mini-batches from the replay memory and applies the DDQN mechanism to compute stable target Q-values. Specifically, the action is selected using the current Q-network, but its value is estimated using the target network. The algorithm then computes the augmented Lagrangian loss, which reflects both reward maximization and constraint handling, and updates the model parameters through gradient descent. The target network is periodically synchronized with the main Q-network to maintain training stability. This process is repeated over multiple episodes and time steps until convergence, ultimately yielding a trained model capable of making real-time CPM decisions that adhere to both performance and QoS constraints.

At each time step of the execution phase, the agent observes the current network state and augments it with the current dual variables to form the input for decision-making. Using the optimized model parameters from training, the agent makes a CPM decision that aims to optimize performance while meeting QoS requirements. Every $T_0$ steps, the algorithm updates the dual variables based on the average constraint violations observed over the last epoch according to (\ref{eq:AugmentedLambdaUpdateIteration}). This update mechanism allows the agent to dynamically adjust its behavior in response to evolving constraint pressures, maintaining a balance between primary objectives and QoS guarantees throughout deployment. The training and execution methodologies for QaSAL-CPM utilizing the DDQN with experience replay are summarized in Algorithms~\ref{alg:Algorithm1} and~\ref{alg:Algorithm2}, respectively. 

\begin{figure}
    \begin{algorithm}[H]
        \caption{Training phase of QaSAL-CPM with DDQN.}
        \label{alg:Algorithm1}
            \textbf{Input:} Number of training episodes $N$, number of time steps $T$, QoS constraints $\mathbf{c}$, target network update frequency $\tau$, batch size $B$, replay memory capacity $M$, primal learning rate $\eta_{\boldsymbol{\Tilde{\theta}}}\in(0,1]$, discount factor $\gamma\in(0,1]$. \\
            \textbf{Output:} Optimal model parameters $\boldsymbol{\Tilde{\theta}}^*$. \\
            \begin{algorithmic}[1]
            \setstretch{1.1}
            \STATE Initialize: \\ 
                \quad Model parameters $\boldsymbol{\Tilde{\theta}}_0$. \\
                \quad Target parameters $\boldsymbol{\Tilde{\theta}}^{\text{target}}\gets\boldsymbol{\Tilde{\theta}}_0$. \\
                \quad Experience replay memory $\mathpzc{D}\gets\varnothing$. \\
                \STATE \textbf{for} $n=0,1,\dots,N-1$ \textbf{do}
            \STATE \quad Observe the initial network state $\mathbf{S}_0$. \\
            \STATE \quad Randomly sample dual variables $\boldsymbol{\lambda}_n=\left\{\lambda_{i,n}\sim p_{\lambda}\right\}_{i=1}^{m-1}$. \\
            \setstretch{1.2}
            \STATE \quad \textbf{for} $t=0,1,\dots,T-1$ \textbf{do} \\
                \quad \scalebox{0.85}{a:} Augment network state $\mathbf{\Tilde{S}}_t=(\mathbf{S}_t,\boldsymbol{\lambda}_n)$. \\
                \quad \scalebox{0.85}{b:} Generate CPM decision $\mathbf{a}_t=\mathbf{a}(\mathbf{\Tilde{S}}_t; \boldsymbol{\Tilde{\theta}}_n)$. \\
                \quad \scalebox{0.85}{c:} Calculate main reward $r_t=f_0(\mathbf{\Tilde{S}}_t, \mathbf{a}_t)$. \\
                \quad \scalebox{0.85}{d:} Calculate violations $v_t=\sum_{i=1}^{m-1}{\lambda_{i,n}\left(f_i(\mathbf{\Tilde{S}}_t, \mathbf{a}_t)-\text{c}_i\right)}$. \\
                \quad \scalebox{0.85}{e:} Observe and augment next state $\mathbf{\Tilde{S}_{t+1}}=\left(\mathbf{S}_{t+1},\boldsymbol{\lambda}_n\right)$. \\
                \quad \scalebox{0.85}{f:} Store transition $\left(\mathbf{\Tilde{S}}_t, \mathbf{a}_t,r_t,v_t,\mathbf{\Tilde{S}}_{t+1}\right)$ in $\mathpzc{D}$. \\
                \quad \scalebox{0.85}{g:} Sample minibatch $\left\{\left(\mathbf{\Tilde{S}}_j, \mathbf{a}_j,r_j,v_j,\mathbf{\Tilde{S}}_{j+1}\right)\right\}_{j=1}^{B}$ from $\mathpzc{D}$. \\
                \quad \scalebox{0.85}{h:} Calculate $\left\{y_j=r_j+\gamma\max_{\mathbf{a}^{\prime}}{Q(\mathbf{\Tilde{S}}_{j+1},\mathbf{a^{\prime}}\mid\boldsymbol{\Tilde{\theta}}_n^{\text{target}})}\right\}_{j=1}^{B}$. \\
                \quad \scalebox{0.85}{i:} Compute the augmented Lagrangian loss: \\
                    \setstretch{1.5}
                    \quad \quad \quad $L(\boldsymbol{\Tilde{\theta}}_n)=\frac{1}{B}\sum_{j=1}^{B}{\left(y_j + v_j -Q(\mathbf{\Tilde{S}}_j,\mathbf{a}_j\mid\boldsymbol{\Tilde{\theta}}_n)\right)^2}$. \\ 
                \quad \scalebox{0.85}{j:} Update model parameters via gradient descent: \\
                    \quad \quad \quad $\boldsymbol{\Tilde{\theta}}_{n+1} \gets \boldsymbol{\Tilde{\theta}}_n - \eta_{\boldsymbol{\Tilde{\theta}}}\nabla_{\boldsymbol{\Tilde{\theta}}_n}{L(\boldsymbol{\Tilde{\theta}}_n)}$. \\
                \quad \scalebox{0.85}{k:} Update $\boldsymbol{\Tilde{\theta}}^{\text{target}}\gets\boldsymbol{\Tilde{\theta}}_{n+1}$ every $\tau$ steps. \\
            \STATE \quad \textbf{end for} \\
            \setstretch{1} 
            \STATE \textbf{end for} \\
            \setstretch{1.5}
            \STATE $\boldsymbol{\Tilde{\theta}}^*\gets \boldsymbol{\Tilde{\theta}}_N$
        \end{algorithmic}
    \end{algorithm}
    
    \begin{algorithm}[H]
        \caption{Execution phase of QaSAL-CPM.}
        \label{alg:Algorithm2}
        \textbf{Input:} Optimal model parameters $\boldsymbol{\Tilde{\theta}}^*$, dual variables update rates $\boldsymbol{\eta}_{\lambda}$, QoS constraints $\mathbf{c}$, epoch duration $T_0$. \\
        \textbf{Output:} Sequence of CPM decisions $\{\mathbf{a}_t; t=0,1,...\}$. \\
        \begin{algorithmic}[1]
            \setstretch{1.1}
            \STATE Initialize: $\boldsymbol{\lambda}_0\gets\boldsymbol{0}$, $k\gets 0$.
            \STATE \textbf{for} $t=0,1,...$ \textbf{do} \\
            \quad \scalebox{0.85}{a:} Observe and augment the state $\mathbf{\Tilde{S}_{t}}=\left(\mathbf{S}_{t},\boldsymbol{\lambda}_k\right)$. \\
            \quad \scalebox{0.85}{b:} Generate CPM decision $\mathbf{a}_t=\mathbf{a}(\mathbf{\Tilde{S}}_t; \boldsymbol{\Tilde{\theta}}^*)$. \\
            
            \quad \scalebox{0.85}{c:} \textbf{if} $t+1\mod T_0=0$ \textbf{then} \\
            \quad \quad \quad Update the dual variables $\left\{\lambda_i\right\}_{i=1}^{m-1}$ as: \\
            \setstretch{1.5}
                \quad \quad \quad \quad $\lambda_{i,k+1} = \left[\lambda_{i,k}-\frac{\eta_{\lambda_i}}{T_0}\sum_{t=kT_0}^{(k+1)T_0-1} \left(f_i(\mathbf{\Tilde{S}}_t, \mathbf{a}_t)-\text{c}_i\right)\right]^+$. \\
                \quad \quad \quad \quad $k\gets k+1$. \\
            \quad \scalebox{0.85}{d:} \textbf{end if}    
            \STATE \textbf{end for}
            \setstretch{1} 
        \end{algorithmic}
    \end{algorithm}
\end{figure}

\section{Performance Evaluation and Simulation Results}
\label{sec:QaSAL-CPM-Coexistence}

In this section, we evaluate the proposed CPM algorithms in a realistic coexistence scenario involving 5G NR-U and Wi-Fi systems sharing an unlicensed spectrum. 
We implemented a Python simulation environment based on {\em SimPy} that models the MAC-layer behavior of 5G NR-U and Wi-Fi under saturated traffic conditions. The simulated wireless environment is inherently dynamic and competitive due to the heterogeneous nature of the technologies involved. While 5G NR-U relies on scheduled transmissions with LBT, Wi-Fi uses a distributed contention-based access mechanism (i.e., CSMA/CA). These differences can lead to severe performance disparities and unpredictable behavior when the medium becomes congested.
To investigate the behavior of the proposed methods, we first apply the multi-objective~RL based algorithm from Section~\ref{sec:MORL-CPM}, which we call MORL-CPM, to a representative coexistence setting to demonstrate its limitations in handling competing objectives under hard QoS constraints. Subsequently, we apply the proposed QaSAL-CPM algorithm from Section~\ref{sec:QaSAL-CPM} across two distinct scenarios:
\begin{itemize}
    \item \textit{Scenario 1:} A single gNB PC1 transmitter coexists with a fixed set of 25 PC3 transmitters from both NR-U and Wi-Fi networks.
    \item \textit{Scenario 2:} The gNB PC1 transmitter coexists with a symmetric and varying number of PC3 transmitters from both gNB and AP sides, ranging from 0 to 50.
\end{itemize}

\subsection{Performance Metrics}

We focus on two key performance metrics to evaluate the effectiveness of CPM: \textit{medium access delay} and \textit{airtime fairness}. Medium access delay is a critical factor for ensuring low-latency communication, particularly for high-priority traffic, while Jain’s fairness index provides a quantitative measure of how equitably resources are distributed among competing nodes. The two metrics are elaborated below.

\subsubsection{Medium Access Delay}

In coexistence scenarios where 5G NR-U and Wi-Fi share unlicensed spectrum, medium access delay plays a crucial role in ensuring Quality of Service (QoS), particularly for high-priority traffic such as
URLLC, which demands stringent latency guarantees, often requiring end-to-end delays as low as $1$~ms with high reliability. Any excessive delay in accessing the channel or transmitting packets can lead to QoS degradation, packet drops, or violations of reliability constraints, making medium access delay a critical metric in CPM. By dynamically adjusting the contention parameters, the medium access delay can be controlled to ensure that high-priority URLLC traffic meets its latency requirements while maintaining fairness for other coexisting users. 

\begin{figure}
\centering
    \includegraphics[scale=0.265]{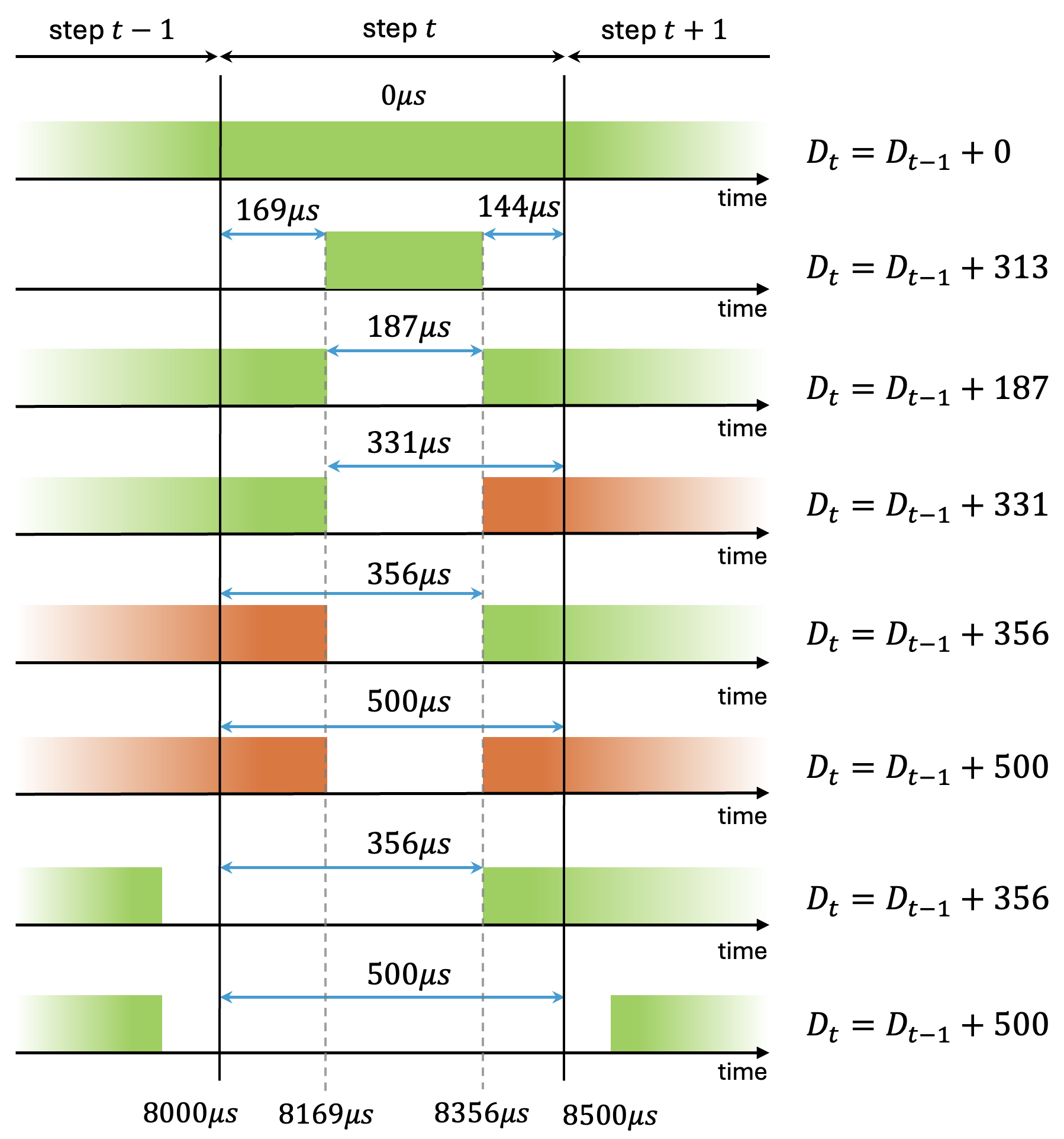}
    \caption{Illustration of medium access delay calculation across time steps. Each row represents a different scenario where transmissions (green) and collisions (orange) occur within a given time step.}
    \label{fig:StepDelay}
    \vspace*{-0.1in}
\end{figure}

In this paper, we define medium access delay as the time interval between the completion of a node's successful transmission and the start of its next successful transmission. This delay includes backoff delay, which accounts for the time spent in the exponential backoff process before attempting transmission, and contention delay, which encompasses the total time a node waits due to channel occupancy by other devices. Additionally, when a collision occurs, the node must restart the contention process, further increasing the delay before the next successful transmission. Fig.~\ref{fig:StepDelay} illustrates the calculation of medium access delay across the learning time steps. The cumulative medium access delay, $D_t$, is experienced by a node up to the end of time step $t$, accounting for the delays incurred in successful transmissions and collisions. 

In an NR-U/Wi-Fi coexistence scenario the delay experienced by a high-priority PC1 transmitter is strongly influenced by the activity of other coexisting nodes, particularly lower-priority transmitters like those in PC3. These nodes often operate with fixed Maximum Channel Occupancy Time (MCOT), such as $8$~ms for PC3, which means that when a PC1 transmitter has to wait for the channel, its delay tends to increase in discrete steps. For instance, the delay could jump to 8~ms, 16~ms, or more, depending on how many consecutive MCOT periods it is blocked. As a result, the delay values do not vary smoothly but appear in distinct levels. This makes it difficult to access performance based on a single delay value, since it might be unusually high due to transient interference or contention.To address this, we use a \textit{smoothed} medium access delay, $\overline{D}_t$, which averages the last few delays (e.g., the last five) observed for the transmitter. This helps reduce the sensitivity to transient spikes caused by contention or collision, and provides a more consistent reflection of ongoing latency performance. As a result, the learning algorithm can make more informed and stable decisions when optimizing coexistence under QoS constraints.

\subsubsection{Airtime Fairness}

While reducing medium access delay is critical for high-priority traffic, it must be balanced with fair resource allocation between PC1 and PC3 transmitters from both the NR-U and Wi-Fi networks. An overly aggressive optimization for PC1 traffic could starve lower-priority traffic, leading to unfair spectrum access and potential performance degradation. To prevent this, fairness should be incorporated as a key metric in CPM optimization. Jain’s Fairness Index (JFI) is a widely used metric to assess the fairness of resource allocation among competing entities. Considering the PC1 and PC3 transmitters as these entities, we define JFI as follows:
\begin{equation}
    \text{JFI}=\frac{\left(\text{Airtime}_{\text{ PC1}}+\text{Airtime}_{\text{ PC3}}\right)^2}{2\left(\text{Airtime}_{\text{ PC1}}^2+\text{Airtime}_{\text{ PC3}}^2\right)} .
\end{equation}
The JFI value ranges from $0.5$ to $1$, with values closer to $1$ indicating a more equitable distribution of airtime. This ensures that neither PC1 nor PC3 traffic classes disproportionately dominates spectrum access. 

This work initially examines the CPM problem in a general form with multiple objective functions and introduces algorithms to address it. As a case study, the focus is then narrowed to optimizing the JFI between PC1 and PC3 traffic classes, while ensuring that the medium access delay of PC1 transmitter remains below a specified threshold.

\subsection{System Setting}

Next, we detail the implementation setup used for evaluating the proposed CPM framework in the context of 5G NR-U and Wi-Fi coexistence. The network state $\mathbf{S}_t$ is designed to comprehensively capture the dynamic environment of 5G NR-U and Wi-Fi coexistence on an unlicensed spectrum. It encodes critical metrics that influence decision-making, including network performance and resource utilization such as the average and smoothed medium access delay of PC1 transmitter, the collision rates, channel utilization airtime ratio, rate of QoS violations, and JFI. Additionally, it tracks trends in delay variation and short-term collision statistics to provide insights into ongoing network conditions.

The CPM decisions $\textbf{a}_t=\textbf{a}(\boldsymbol{\textbf{S}_t})$ are selected from the set $\{0,1,\ldots,6\}$. The maximum contention window (CW) size is calculated as $\text{CW}_{\text{max, PC}i}=2^{\text{a}_{i,t}+b}-1$, where $\text{a}_{i,t}$ is the CPM decision for priority class~$i$, and $b=0$ and $4$ for PC1 and PC3 traffic, respectively. This design directly influences the backoff timing, which affects each transmitter's access to the channel. Moreover, the objectives are designed as $f_0(\mathbf{S}_t, \mathbf{a}(\mathbf{S}_t))=\mathrm{JFI}_t$ and $f_1(\mathbf{S}_t, \mathbf{a}(\mathbf{S}_t))=\overline{D}_{\text{PC}1,t}$, denoting the JFI among PC1 and PC3 transmitters and the smoothed medium access delay of PC1 transmitter at time $t$, respectively. We define a constraint for medium access delay of PC1 in terms of a threshold $D_{\text{th}}$. 

In MORL-CPM, the agent jointly optimizes medium access delay and airtime fairness through a weighted objective function. In contrast, QaSAL-CPM maximizes the airtime fairness between two traffic classes while ensuring that the medium access delay experienced by PC1 transmitter does not exceed a predefined threshold. To learn these policies, we adopt the DDQN with experience replay proposed in Section~\ref{sec:DRL-QASAL-CPM}. Each training agent receives an augmented state comprising the system observations and, in the case of QaSAL-CPM, the dual variables associated with delay constraints. Learning is conducted in episodes, each representing a fixed-duration coexistence period.

\begin{figure*}
\begin{multicols}{2}
\centering
        \includegraphics[width=0.95\linewidth]{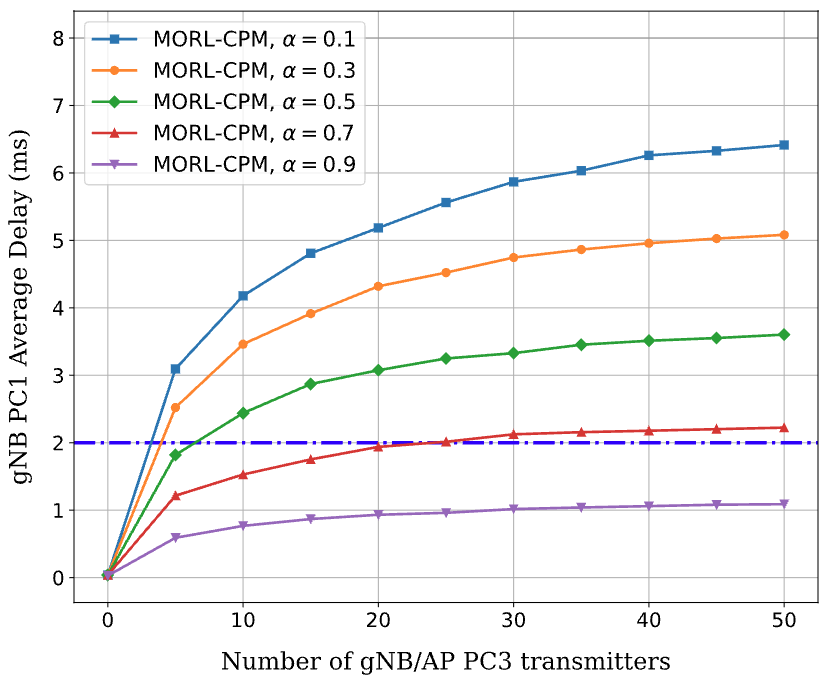}
        \label{fig:MORL-CPM_Delay_PC1}
    \par
	\includegraphics[width=0.95\linewidth]{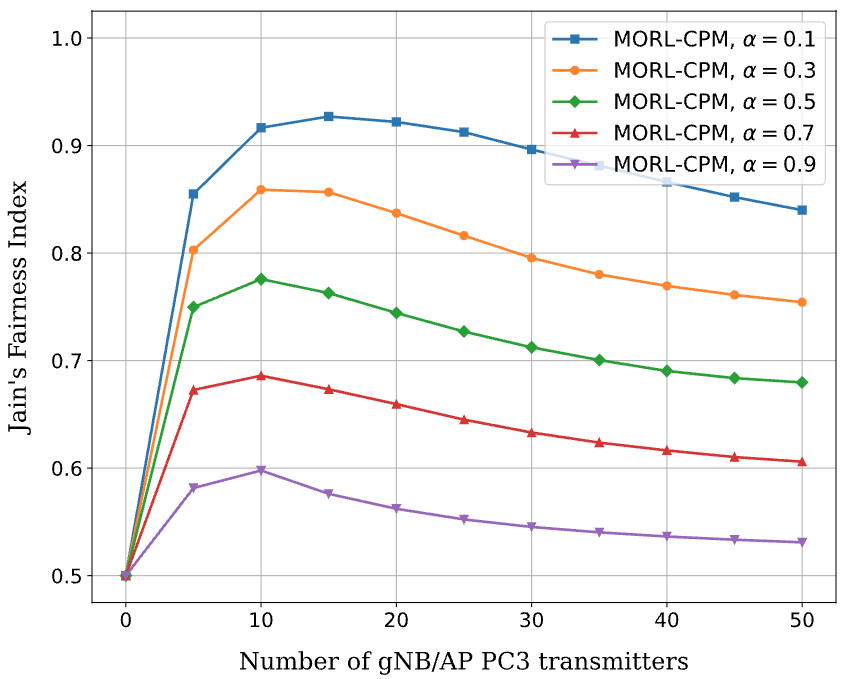}
        \label{fig:MORL-CPM_JFI}
    \par
\end{multicols}
\vspace*{-0.3in}
\caption{Average medium access delay of gNB PC1, and fairness between PC1 and PC3 transmitters, under the MORL-CPM algorithm across various delay thresholds in Scenario~2. The dashed line represents the QoS threshold $D_{\text{th}}=2~\text{ms}.$}
\label{fig:MORL_Performance_Metrics_3_Class}
\end{figure*}

\subsection{Problem Formulation}

We first solve the CPM for coexistence of 5G NR-U and Wi-Fi through the MORL-CPM algorithm. Utilizing the linear scalarization approach, we define the weight vector as $\boldsymbol{w}=\{1-\alpha, \alpha\}$ according to (\ref{eq:Linear-Scalarized-Objective}). The trade-off parameter $\alpha\in[0,1]$ balances the relative importance of the objectives. Higher values of $\alpha$ put more emphasize on reducing $\overline{D}_{\text{PC}1,t}$, whereas lower values put more weight on JFI. We normalize  $\overline{D}_{\text{PC}1,t}$ by $\text{D}_{\text{max}}$ to normalize the delay term to $[0, 1]$. This is critical for combining it with JFI, which is inherently bounded in the same range. Therefore, the \textit{multi-objective CPM} problem in (\ref{eq:Parameterized-Linear-Scalarized-CPM}) can be formulated as
\begin{equation}
    \max_{\{\textbf{a}_t\}_{t=0}^{T-1}}{\quad \frac{1}{T}\sum_{t=0}^{T-1}{\left\{(1-\alpha) \mathrm{JFI}_t+\alpha \left(1-\frac{\overline{D}_{\text{PC}1,t}}{\text{D}_{\text{max}}}\right)\right\}}}.
    \label{eq:Parameterized-Linear-Scalarized-CPM-Coexistence}
\end{equation}

In contrast, the formulation for QaSAL-CPM introduces the \textit{parameterized constrained CPM} problem as follows:
\begin{subequations}
    \begin{align}
        \max_{\{\textbf{a}_t\}_{t=0}^{T-1}} \quad &
            \frac{1}{T}\sum_{t=0}^{T-1} \mathrm{JFI}_t,
            \label{eq:ParameterizedCPMCoexistence:Objective} \\
        \text{s.t.} \quad &
            \frac{1}{T}\sum_{t=0}^{T-1} \overline{D}_{\text{PC}1,t}\leq D_{\text{th}} 
            \label{eq:ParameterizedCPMCoexistence:Constraints}
    \end{align}
    \label{eq:ParameterizedCPMCoexistence}
\end{subequations}
This constrained formulation is solved via state-augmented~RL, where the Lagrange multipliers associated with delay constraints are updated online and included in the agent's observation space, enabling real-time adaptation to varying traffic dynamics. 

\subsection{Congestion-Aware Dual Adjustment and Reward Design}

To make the agent more responsive to constraint violations without being overly conservative in low-load conditions, we apply a two-part strategy. First, the dual update step size $\eta_{\lambda}$ in (\ref{eq:AugmentedLambdaUpdateIteration}) is adjusted based on the violation rate, which is the percentage of time steps in which the medium access delay exceeds the threshold:
\begin{equation}
    \eta_{\lambda,k} = \eta_{\lambda,\text{min}} + \left(\eta_{\lambda,\text{max}} - \eta_{\lambda,\text{min}}\right) \rho_{\text{violation},k} ,
    \label{eq:eta_lambda_adjustment}
\end{equation}
where $\eta_{\lambda,\text{min}}$ and $\eta_{\lambda,\text{max}}$ are minimum and maximum learning rates for dual updates, and $\rho_{\text{violation},k}$ is the ratio of time steps with delay violations over the update window size, $T_0$. This makes $\lambda$ update more aggressively when violations are frequent, and less aggressively when the constraint is approximately satisfied. Second, the reward is computed as 
\begin{equation}
    r_t = \text{JFI}_t - \beta \max\left(0, \frac{D_{\text{th}}-\overline{D}_{\text{PC}1,t}}{D_{\text{th}}}\right) ,
    \label{eq:reward_shaping}
\end{equation}
where $\beta$ is a scaling factor that controls the strength of the penalty applied when the delay falls below the threshold, which encourages the agent to utilize the delay budget more effectively in low-congestion scenarios. In our simulations, $\beta$ was set to a value proportional to the delay threshold; in particular, we used
$\beta = 1, 1.5, 2$ for $D_{\text{th}} = 1$~ms, $2$~ms, and $3$~ms, respectively.

\begin{table}
    \centering
    \caption{Simulation Setup}
    \renewcommand{\arraystretch}{1.05}
    \begin{tabular}{ c  c }
        \Xhline{2\arrayrulewidth}
        Parameter & Value \\ 
        \hline
        Interaction time & $10,000$~episodes \\
        Episode duration & $500$~steps \\
        Step duration & $2.5~\text{ms}$ \\
        Discount factor & $0.99$ \\ 
        Replay buffer size & 100,000 \\
        Range of $\epsilon$ & 1.0 to 0.01 \\ 
        DQN learning rate & $10^{-5}$ \\
        Batch size & 16 \\
        Hidden layers & $128 \times 128 \times 128$\\
        $\lambda_{\text{max}}, T_0, \eta_{\lambda}$ & 5.0, 5, [0.01, 0.2]\\
        \Xhline{2\arrayrulewidth}
    \end{tabular}
    \label{tab:DQN_param}
    \vspace*{-0.1in}
\end{table}

\subsection{Simulation Results and Analysis}
\label{sec:SimulationResults}

This subsection presents the performance of the proposed methods under different traffic and interference scenarios. We first apply MORL-CPM in Scenario~2 and evaluate its performance. Next, we compare QaSAL-CPM with the Primal-Dual approach in Scenario~1. Finally, we analyze the effectiveness of QaSAL-CPM under varying delay thresholds and network congestion levels. The hyperparameters used for simulation are summarized in Table~\ref{tab:DQN_param}. The step duration is selected to be large enough to include several transmission attempts to enable accurate calculation of the medium access delay. 

\begin{figure*}
\begin{multicols}{2}
\centering
    \includegraphics[width=0.95\linewidth]{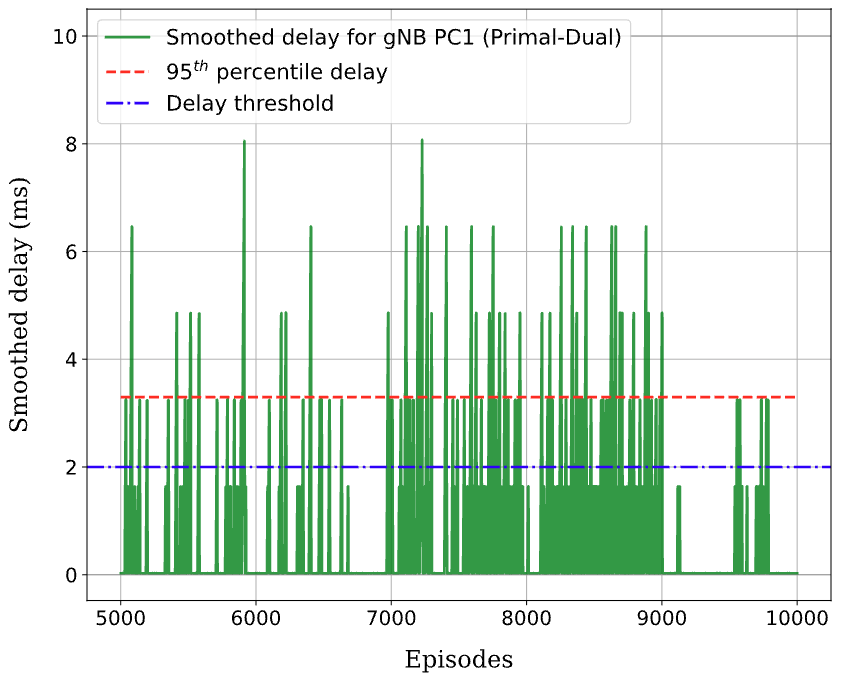}
        \label{fig:PrimalDual_Execution_Delay_PC1}
    \par
	\includegraphics[width=0.95\linewidth]{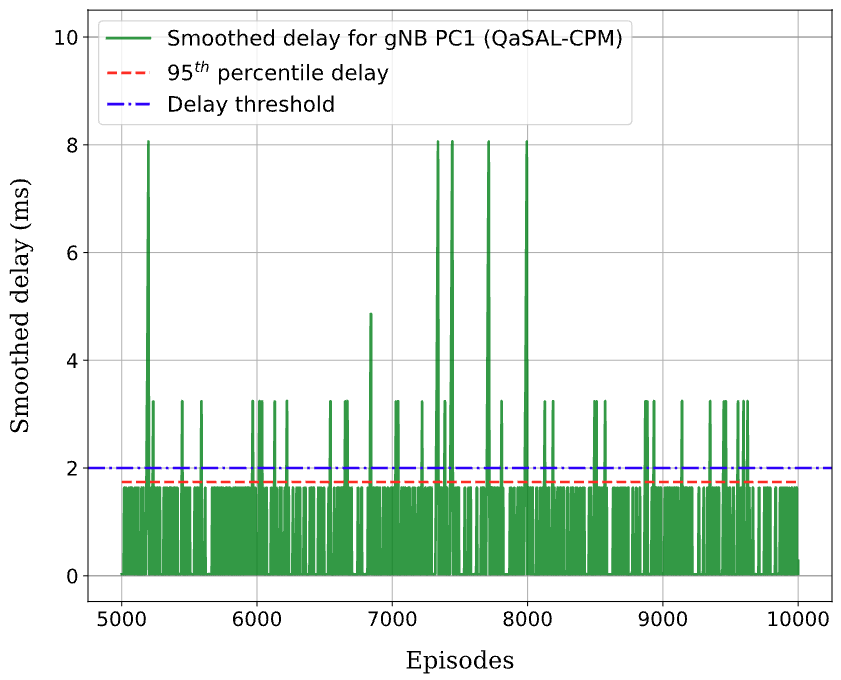}
        \label{fig:QaSAL_Execution_Delay_PC1}
    \par
\end{multicols}
\vspace*{-0.3in}
\caption{Evolution of smoothed medium access delay for gNB PC1 with Primal-Dual and QaSAL-CPM algorithms in Scenario~1 ($D_{\text{th,PC1}}=2~\text{ms}$).}
\label{fig:Execution_Delay_PC1_2_Class}
\end{figure*}

\subsubsection{MORL-CPM}

Fig.~\ref{fig:MORL_Performance_Metrics_3_Class} illustrates how the trade-off coefficient $\alpha$ (see eq.~\ref{eq:Parameterized-Linear-Scalarized-CPM-Coexistence}) in the MORL-CPM framework affects the average medium access delay experienced by a PC1 transmitter and JFI between PC1 and PC3 transmitters from both networks, respectively. As the number of PC3 transmitters increases, the delay performance degrades across all configurations; however, the degree of degradation depends significantly on the choice of $\alpha$. Notably, lower values of $\alpha$, which place greater emphasis on the fairness (as the secondary objective), result in higher delays for the PC1 transmitter—clearly exceeding the delay threshold of 2~ms. Even moderate scalarization settings (e.g., $\alpha=0.5$) fail to consistently enforce the QoS constraint as the network becomes more congested. Only high scalarization values (e.g., $\alpha=0.7$ and $\alpha=0.9$), which strongly prioritize delay minimization for PC1, maintain the delay below or near the required threshold. 

These results highlight a key limitation of scalarized multi-objective based optimization in constrained environments: there is no guarantee of strict QoS satisfaction for high-priority traffic, especially under varying network conditions. This limitation motivates the need for more explicit constraint-handling strategies.

\begin{figure*}
\begin{multicols}{2}
\centering
    \includegraphics[width=0.95\linewidth]{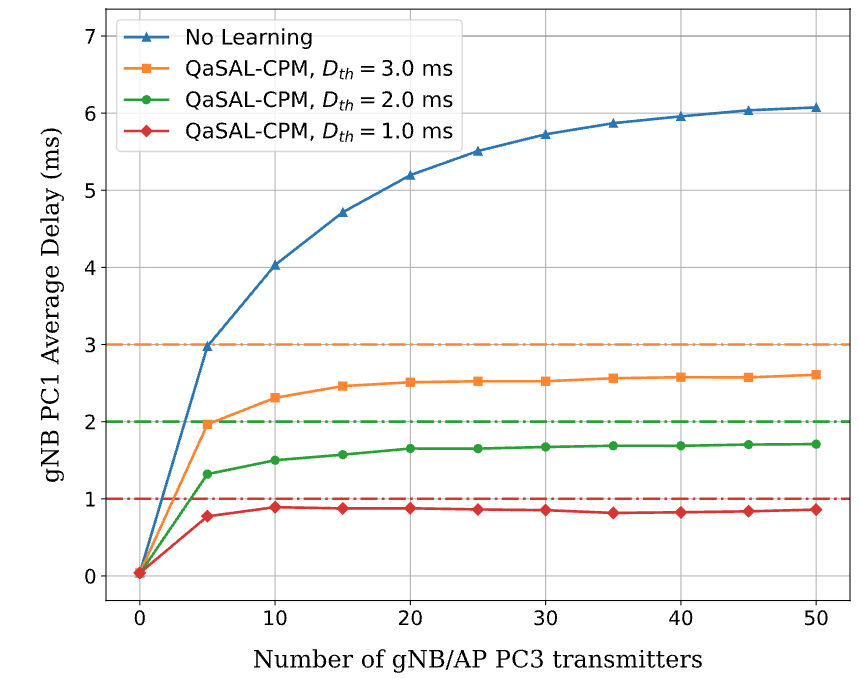}
        \label{fig:QaSAL_Delay_PC1_3_Class}
        \par
    \includegraphics[width=0.95\linewidth]{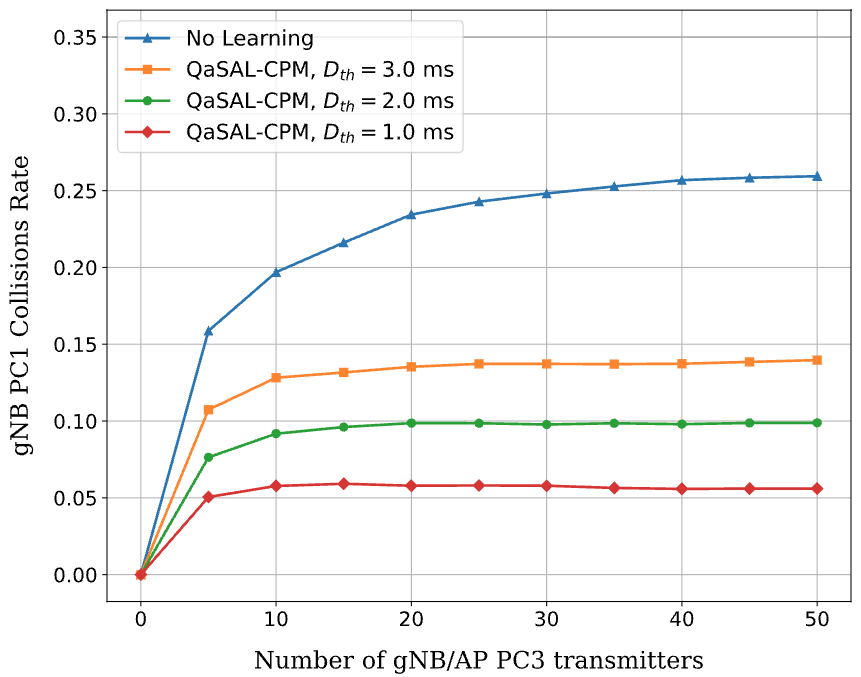}
        \label{fig:QaSAL_Collision_PC1_3_Class}
        \par
\end{multicols}
\begin{multicols}{2}
\centering
    \includegraphics[width=0.95\linewidth]{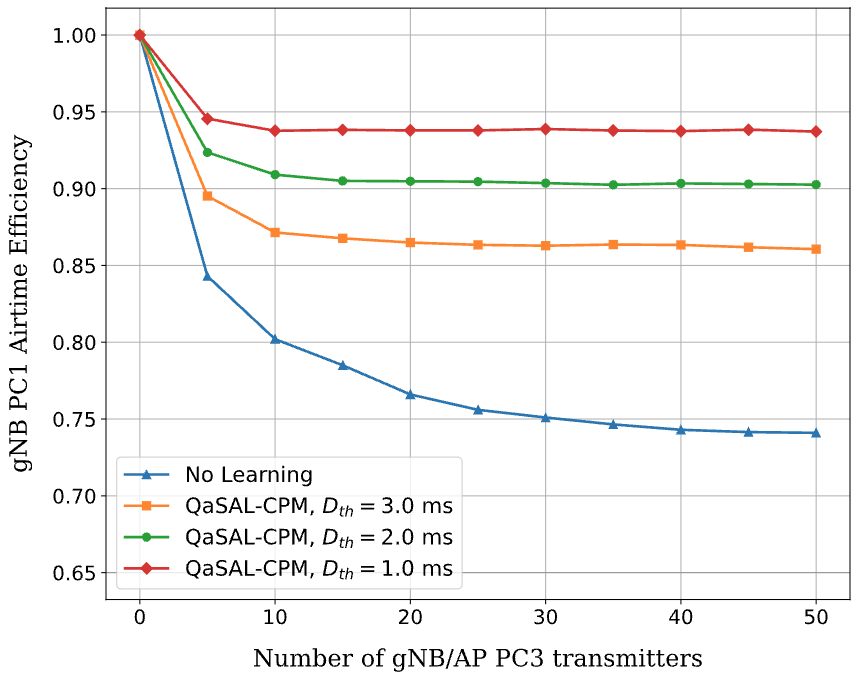}
        \label{fig:QaSAL_Efficiency_PC1_3_Class}
        \par
    \includegraphics[width=0.95\linewidth]{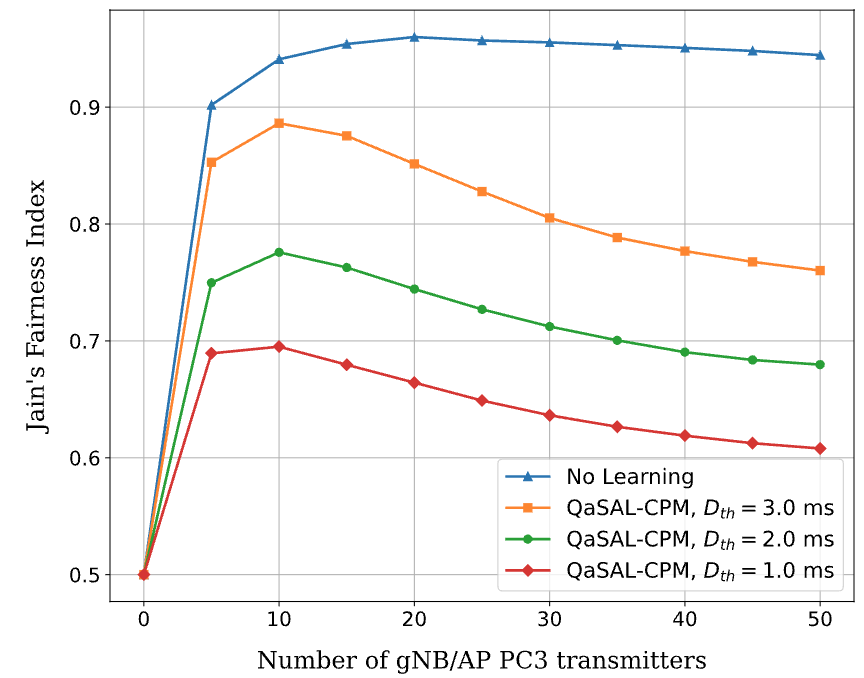}
        \label{fig:QaSAL_JFI_PC1_3_Class}
        \par
\end{multicols}
\vspace*{-0.1in}
\caption{Average medium access delay, collision rate, and airtime efficiency of gNB PC1, along with the fairness between PC1 and PC3 transmitters, under the QaSAL algorithm across various delay thresholds in Scenario~2. The dashed lines represents the QoS thresholds $D_{\text{th}}=\text{1.0, 2.0, and 3.0~ms}.$}
\label{fig:QaSAL_Performance_Metrics_3_Class}
\end{figure*}

\subsubsection{QaSAL-CPM}

We compare the performance of the primal-dual approach (i.e., no state augmentation) and the proposed QaSAL method under Scenario~1. Fig.~\ref{fig:Execution_Delay_PC1_2_Class} illustrates the evolution of the smoothed medium access delay for gNB PC1 for the Primal-Dual and QaSAL-CPM algorithms, respectively, over 5000 execution episodes, each of duration $1.25$~s. While the Primal-Dual method satisfies the delay constraint on average, it frequently experiences constraint violations, indicating weaker adherence to real-time QoS requirements. The horizontal, dashed red lines in Fig.~\ref{fig:Execution_Delay_PC1_2_Class} indicate the $95$-th percentile smoothed delays over the execution duration, which is a commonly used performance metric for URLLC traffic. For the Primal-Dual method, the $95$-th percentile delay well exceeds the $2$~ms delay threshold, whereas it lies below the threshold for QaSAL-CPM. 


In contrast, QaSAL-CPM leverages the state-augmentation approach to maintain the delay consistently below the specified threshold which results in significantly better compliance with QoS constraints. The delay dynamics in QaSAL-CPM are directly influenced by the evolution of the dual variable associated with the delay constraint, as defined in equation (\ref{eq:ParameterizedCPMCoexistence:Constraints}). Compared to the Primal-Dual approach, QaSAL-CPM demonstrates a more adaptive and responsive adjustment of the dual variable by spiking in response to constraint violations and decaying as the constraint becomes satisfied and enabling more reliable and timely QoS management.


Fig.~\ref{fig:QaSAL_Performance_Metrics_3_Class} presents the average medium access delay experienced by the gNB PC1 transmitter, collision rate and airtime efficiency of gNB PC1, and the fairness across PC1 and PC3 transmitters from both networks, under different delay threshold configurations in Scenario~2. The results clearly show that the QaSAL-CPM algorithm effectively maintains the average delay close to the desired delay levels, whereas more relaxed thresholds lead to higher allowable delays, showing the algorithm's ability to adapt accordingly. The performance under the ``No Learning'' baseline exhibits increasing delay as the number of contending transmitters grows, validating the importance of a learning-based dynamic adaptation mechanism like QaSAL in managing coexistence. Notably, the algorithm successfully stabilizes delay performance even as the network becomes increasingly congested. The fairness performance under QaSAL-CPM is strongly influenced by the delay threshold configuration. Tighter thresholds, while reducing delay, tend to lead to lower fairness due to the algorithm prioritizing the delay requirement  over the airtime distribution. Conversely, more relaxed delay constraints allow for improved fairness, as seen with the $2.0$~ms and $3.0$~ms delay threshold configurations. 
Note that when the delay threshold is higher, the agent keeps the delay further below it to avoid violations caused by the added congestion that comes from trying to improve fairness. Overall, these results highlight the inherent trade-off between delay and fairness in constrained multi-objective optimization and demonstrate that QaSAL-CPM can flexibly manage these trade-offs through its tunable delay constraints. 

In our experiments, we assumed a worst-case scenario in which the PC1 transmitter's buffer is always full. In a more realistic scenario, there may be multiple PC1 transmitters, each of which transmits intermittently, allowing more airtime for PC3 traffic. In this paper, QaSAL-CPM controls only one parameter, the CW size. More fine-grained QoS control may be achievable by extending QaSAL-CPM to manage both CW and another parameter, such as AIFS number (see~\cite{Hirzallah:2019}). The delay performance could be improved further by adopting an enhanced version of LBT that employs collision resolution, e.g.,~\cite{Loginov:2021}. We plan to investigate these and related issues in our future work. 
\section{Conclusions}
\label{sec:Conclusion}

We investigated the application of reinforcement learning to CPM in heterogeneous wireless networks, with a specific focus on the coexistence of 5G NR-U and Wi-Fi. 
We proposed a novel state-augmentated, constrained~RL solution called QaSAL-CPM that directly embeds dual variables into the state space. This design enables the learning agent to adapt its policy in real time with improved stability and convergence, relative to
constrained~RL based on conventional primal-dual methods and multi-objective RL. 

We validated our approach through extensive simulations under multiple realistic coexistence scenarios. The simulation results demonstrated that QaSAL-CPM not only satisfies the imposed QoS constraints but also provides smoother training dynamics and improved performance consistency compared to the primal-dual method. Overall, the proposed framework provides a flexible and generalizable approach for managing coexistence parameters in dynamic and complex wireless environments, offering a promising direction for future research.



\bibliographystyle{IEEEtran}
\bibliography{IEEEabrv,main}

\end{document}